\documentclass[%
aip,
amsmath,amssymb,
floatfix,
]{revtex4-1}

\usepackage{graphicx}% Include figure files
\usepackage{placeins}
\usepackage{dcolumn}% Align table columns on decimal point
\usepackage{bm}% bold math
\usepackage[utf8]{inputenc}
\usepackage[T1]{fontenc}
\usepackage{mathptmx}
\usepackage{xcolor}
\begin{document}
\title[]{Rotating vortices in two-dimensional inhomogeneous strongly coupled dusty plasmas:  shear and spiral-density waves}
%\title[]{Fluid simulation  of two-dimensional vortex dynamics in inhomogeneous strongly coupled dusty plasmas}

\author{Vikram S. Dharodi}%
\email{dharodiv@msu.edu}
\affiliation{Mechanical Engineering, Michigan State University, East Lansing, Michigan 48824, USA.}
	\date{\today}% It is always \today, today,
	\begin{abstract}
%  	\large
Dusty plasma experiments can be performed quite easily in strong coupling  regime. In our previous work~[Phys. Plasmas 21, 073705 (2014)], we numerically explored such plasmas with constant density and observed the transverse shear (TS) waves from a rotating vortex. Laboratory dusty plasmas are good examples of homogeneous plasmas however heterogeneity (e.g. density, temperature, charge) may be due to the existence of voids, different domains with different orientations, presence of external forces like magnetic and/or electric, size/charge imbalance, etc. Here, we examine how the density heterogeneity in dusty plasmas responds to the circularly rotating vortex monopoles, namely smooth and sharp cut-off.  For this purpose we have carried out a series of two-dimensional viscoelastic fluid simulations in the framework of generalized hydrodynamics (GHD) fluid model. The rotating vortices are placed at the interface of two incompressible fluids with different densities. The smooth rotating vortex causes two things: First, the densities are stretched to form the spiral density waves; secondly, the TS waves propagate radially into the surrounding media according to the shear wave speed. We notice that the spiral density arms are distinguishable in the early time while later get smeared out. The sharp rotating vortex creates sharp shear flows which in turn favor the Kelvin-Helmholtz (KH) instability across the interfaces. In such flows for the GHD system, the interplay between the emitted TS waves and the vortices of KH instability distorts the formation of the regular spiral density arms around the rotor. 
\end{abstract}
\maketitle
%%%%%%%%%%%%%%%%%%%%%%%%%%%%%%%%%%%%%%%
%          INTRODUCTION
%%%%%%%%%%%%%%%%%%%%%%%%%%%%%%%%%%%%%%%
\section{Introduction}\label{Introduction}
%  \large
 A typical fluid/turbulent flow contains a wide variety of coherent structures\cite{tur2017coherent}  like rotating monopoles and tripoles\cite{rossi1997quasi,kizner2004tripole}, propagating and  merging dipoles\cite{leweke2016dynamics}. It becomes{\huge } important to understand how such types of structures encounter the density and/or temperature gradient in a medium, which may cause the formation of different kinds of waves like acoustic\cite{lighthill2001waves,rao1990dust,choudhary2016propagation}, shock\cite{grove1989interaction,rotman1991shock,lin2019pressure}, spiral\cite{imao1992characteristics,barkley1992linear}, transverse\cite{peeters1987wigner,vladimirov1997vibrational,schmidt1997longitudinal,wang2001longitudinal,murillo2000critical,liu2003transverse} and fluid instabilities like Kelvin-Helmholtz\cite{drazin_1970,chandrhd}, Rayleigh-Taylor\cite{rayleigh1900investigation,taylor1950instability}. Laboratory dusty plasmas are good examples of homogeneous plasmas however heterogeneity (e.g. density, temperature, charge) may be due to the existence of voids, different domains with different orientations, presence of external forces like magnetic and/or electric, size/charge imbalance, etc.  Our objective here is to understand how the density heterogeneity in a strongly coupled state responds to the coherent structures (here, rotating vortex monopoles). For this purpose we specifically here consider the case  of  two-dimensional (2D) dusty plasmas. The motivational factors that induce us to choose such 2D plasmas are: 2D dusty plasmas are favored in laboratory experiments\cite{thomas1994plasma,thomas1999direct,ichiki2004melting,haralson2015laser,haralson2016temperature,melzer2019finite,kananovich2020experimental} and simulations\cite{lin2019pressure,piel2006laser,feng2016equations,hartmann2019self,lin2020universal}, dusty plasmas can exist in strong coupling state quite easily because of high charged dust particles, and in dusty plasmas a coherent structure can survive for a longer time than in a hydrodynamic fluid (viscosity shows pure damping effect) because the effect of viscosity  gets reduced due to the presence of elasticity \cite{frenkel_kinetic}.  Thus, in dusty plasmas, a long-lived coherent structure can act as a driving force in order to understand its collective response for a long-time duration without much dissipation.  

  Strongly coupled dusty plasmas (SCDPs) behave like viscoelastic mediums cause to support the existence of both incompressible transverse shear and compressible longitudinal modes \cite{nunomura2002dispersion,nunomura2005wave,donko2019molecular}. Here, we have modeled  this medium using well-known phenomenological generalized hydrodynamic (GHD) fluid model which takes into account both  types of modes~\cite{Kaw_Sen_1998, Kaw_2001}.  It should be noted that a fluid model focuses on a situation where the spacial scales in dusty plasmas  are  supposed to be about an order of magnitude larger than the interparticle distance while fails as soon as the grainy structure becomes important. To study the effect of inhomogeneity on exclusive transverse modes and to avoid the possible coupling with longitudinal mode, we consider the incompressible limit of the dusty plasma. Theoretically, the existence of transverse modes in dusty plasma medium has  been predicted in~\cite{peeters1987wigner,vladimirov1997vibrational,wang2001longitudinal}.  Schmidt {\it et al.}  \cite{schmidt1997longitudinal} showed such transverse modes in molecular dynamic simulations.  Nunomura {\it et al.}  \cite{nunomura2000transverse} , Pramanik {\it et al.}   \cite{pramanik2002experimental} and  Pintu {\it et al.}   \cite{bandyopadhyay2008driven}  have also observed transverse modes in dusty plasma experiment.  Using GHD model,  the shear waves  in a nonuniform dusty plasma has been studied in\cite{mishra2000instability,sorasio2003instability}.  Apart from the shear waves, a nonuniform dusty plasma under stretching due to a driving force may cause to support the spiral waves. Recently, Sandeep {\it et al.}  reported the existence of spiral waves in SCDPs at particle level using molecular-dynamics simulations in \cite{kumar2018spiral_pre} and at continuum level using the  GHD model in \cite{kumar2018spiral_pop}. In incompressible fluid simulations, Li {\it et al.}~\cite{li2010nonlinear} found the multi-armed spiral waves in a slowly rotating fluid, driven by a radial unstable temperature gradient.  The existence of spiral waves are observed in various biological systems like retinal spreading depression\cite{gorelova1983spiral}, cardiac muscle\cite{davidenko1992stationary}, and Xenopus oocyte calcium waves~\cite{lechleiter1991spiral}. Apart from the biology, the understanding of spiral waves are important in mathematics, and physics, because the spiral wave formation takes place in nonlinear nonequilibrium systems\cite{kitahata2018mathematical}. A rotating galaxy is a good example of the large scale spirals\cite{elmegreen1981near}. 
  
  In our previous work~\cite{dharodi2014visco}, we explored the constant density dusty plasmas in an incompressible limit of GHD model  and observed the transverse shear (TS) waves due to the shear flow induced from a rotating monopole vortex. In the current work, we use same model where the smooth and sharp cut-off rotating vortices are placed at the interface of two incompressible fluids with different densities.  The smooth rotating vortex at the interface causes two things: first, this rotating vortex convects material from the  higher density part to the lower density part and vice versa which leads to the formation of spiral density waves, and second, locally introduce a shear flow. This shear flow is the source for the shear waves.  These waves propagate into the surrounding media according to the shear wave speed where the waves travel a smaller distance in higher density part than in lower density part in the same time interval.  The sharp rotating vortex creates sharp shear flows which favor the  Kelvin-Helmholtz (KH) instability across their interfaces along with the TS waves. In such flows the interplay between the emitted TS waves and the vortices of KH instability occurs which in turn distorts the formation of the regular spiral density waves around the rotor.  Because of incompressible limit ($ {\nabla}{\cdot}{\vec{v_d}}=0$), to observe the spiral density waves the initial density profile must have a radially varying component to the rotating vortex. Although, here, the density inhomogeneity has been introduced hypothetically (in order to identify the TS waves) for simulation but the presented results can be easily generalized to any typical SCDP experiment having realistic density inhomogeneity $e.g.$, sech type, parabolic type, etc. and by locating the driving vortex at the required place.

This paper is organized as follows. In Section~\ref{ghd_model}, we discuss the incompressible limit of GHD (i-GHD) model used for the SCDPs and  drive an analytical linear wave equation. In Section \ref{num_methodology}, we discuss how the simulations are initialized and analyse the dynamical response of the density heterogeneity to the circularly rotating vortex monopoles, particularly smooth and sharp cut-off. In the subsection~\ref{num_methodology_smooth}, we numerically observe that a smooth rotating vorticity profile emits cylindrical TS waves and also causes to rolling-up the densities to form the spiral density waves.  In the  subsection~\ref{num_methodology_rigid}, for sharp rotating  flows, the interplay between the emitted TS waves and the vortices of KH instability occurs which in turn distorts the formation of the spiral density waves around the rotor. Finally, in Section~\ref{Conclusions}, we discuss our results and offer concluding remarks. 
%%%%%%%%%%%%%%%%%%%%%%%%%%%%%%%%%%%%%%%
% Analytical Description
%%%%%%%%%%%%%%%%%%%%%%%%%%%%%%%%%%%%%%% 
\section{Incompressible generalized hydrodynamics (i-GHD) fluid model} \label{ghd_model}

Generalized hydrodynamic fluid model \cite{Kaw_Sen_1998,diaw2015generalized} is a phenomenological model which is used to study the SCDPs, both analytically as well as numerically \cite{Kaw_Sen_1998, Kaw_2001, dharodi2014visco, dharodi2016sub}. GHD model treat dusty plasma as a viscoelastic fluid in which the coupling strength is proportional to the ratio of ${\eta}/{\tau_m}$~\cite{frenkel_kinetic}, $\eta$ is the shear viscosity coefficient  and $\tau_m$ is the relaxation time parameter. The parameters $\eta$ and $\tau_m$ are supposed to be empirically related to each other \cite{Kaw_Sen_1998, ichimaru1987statistical}. This model supports both the existence of incompressible transverse shear and compressible longitudinal modes. Here, we separate out the compressibility effects altogether and the incompressible limit of the GHD (i-GHD) set of equations has been obtained. In the incompressible limit the Poisson equation is replaced by the quasi-neutrality condition and charge density fluctuations are ignored. The derivation of the reduced equations has been discussed in detail in our earlier papers \cite{dharodi2014visco,dharodi2016sub} along with the procedure of its  numerical implementation and validation. The coupled set of continuity and momentum equations for the dust fluid  can be written as:
%~~~~~~~~~~~~~~~~~~~~~~~~~~~~~~~~~~~~~~~
\begin{equation}\label{eq:continuity}
  \frac{\partial \rho_d}{\partial t} + \nabla \cdot
\left(\rho_d\vec{v}_d\right)=0{,}
  \end{equation}
 %~~~~~~~~~~~~~~~~~~~~~~~~~~~~~~~~~~~~~~~
 \begin{eqnarray}\label{eq:momentum1}
 &&\left[1+{\tau_m}\left(\frac{\partial}{\partial{t}}+{\vec{v}_d}\cdot \nabla\right)\right]\nonumber\\
 && \left[ {{\rho_d}\left(\frac{\partial{\vec{v}_d}}{\partial {t}}+{\vec{v}_d}{\cdot} \nabla{\vec{v}_d}\right)}+{\alpha}{\nabla}{\rho_d}+{\rho_c}\nabla \phi_{d} \right]\nonumber\\
 && =\eta \nabla^2\vec{v}_d{,}
 \end{eqnarray}
%~~~~~~~~~~~~~~~~~~~~~~~~~~~~~~~~~~~~~~~
respectively and the incompressible condition is given as 
%~~~~~~~~~~~~~~~~~~~~~~~~~~~~~~~~~~~~~~~
\begin{equation}\label{eq:incompressible}
 {\nabla}{\cdot}{\vec{v_d}}=0{.}
\end{equation}
%~~~~~~~~~~~~~~~~~~~~~~~~~~~~~~~~~~~~~~~

Here, the variables $\rho_d$ and $\rho_c$, $\vec{v}_d$ and $\phi_d$ are the dust mass density, dust charge density, dust fluid velocity and dust charge potential, respectively.  The pressure is simply a function of density through equation of state and $\alpha$ represents the square of sound speed of the medium. The time, length, velocity and potential are normalised by inverse of dust plasma frequency $\omega^{-1}_{pd} = \left({4\pi (Z_d e)^{2}n_{d0}}/{m_{d0}}\right)^{-1/2}$ and plasma Debye length $\lambda_{d} = \left({K_B T_i}/{4{\pi} {Z_d}{n_{d0}}{e^2}}\right)^{1/2}$, ${\lambda_d}{\omega_{pd}}$ and ${{Z_d}e}/{{K_B}{T_i}}$ respectively. The parameters $m_d$, $T_i$ and $K_B$ are the dust grain mass, ion temperature and Boltzmann constant respectively. $Z_d$ is the charge on each dust grain with no consideration  of charge fluctuation. The number density $n_d$ is normalised by the equilibrium value $n_{d0}$.

%%%%%%%%%%%%%%%%%%%%%%%%%%%%%%%%%%%%%%%
% Analytical Description
%%%%%%%%%%%%%%%%%%%%%%%%%%%%%%%%%%%%%%% 
\subsection{Transverse wave equation} 
\label{anal_description}
If we take the curl of above momentum Eq.~(\ref{eq:momentum1}) and keep the linearized terms only, for the moment, the density is constant $i.e$~ $  {\rho_d}(x,y,t)={\rho_{d}}$, we immediately get the equation 
%~~~~~~~~~~~~~~~~~~~~~~~~~~~~~~~~~~~~~~~
 \begin{equation}\label{}
 \left[1 + \tau_m\frac{\partial}{\partial t}\right]
 \left[{\frac{\partial {\vec \xi} } {\partial t}} \right]
 = {\frac{\eta}{\rho_d}} \nabla^2 {\vec \xi}    
  \end{equation}  
  %~~~~~~~~~~~~~~~~~~~~~~~~~~~~~~~~~~~~~~~
 $ {\vec \xi}={\vec \nabla}{\times}{\vec v_d} $ is the vorticity, normalised with dust plasma frequency. In limit {${\tau_m}{\frac{\partial}{\partial{t}}} \geq 1$}, where the memory effects are strong, we get 
  %~~~~~~~~~~~~~~~~~~~~~~~~~~~~~~~~~~~~~~~
 \begin{equation}\label{eq:wave_eu1}
{\frac{\partial^2 {\vec \xi}} {\partial t^2}}
 = {v_p^2} \nabla^2 {\vec \xi}{.}     
  \end{equation}  
  %~~~~~~~~~~~~~~~~~~~~~~~~~~~~~~~~~~~~~~~
This wave Eq.~(\ref{eq:wave_eu1}) suggests that i-GHD model support the transverse waves moving with phase velocity 
% ${v_p}=\sqrt{{\eta}/{{\rho_d}{\tau_m}}}$. 
   %~~~~~~~~~~~~~~~~~~~~~~~~~~~~~~~~~~~~~~~
 \begin{equation}\label{eq:TSwave_conf}
{v_p}=\sqrt{{\eta}/{{\rho_d}{\tau_m}}}{.}  
 \end{equation}  
 %~~~~~~~~~~~~~~~~~~~~~~~~~~~~~~~~~~~~~~~
 Equation~(\ref{eq:wave_eu1}) also evident that the form of a wave is determined by its source. Let's suppose that we have a line source, the wavefronts will be cylindrical. So, the wave Eq.~(\ref{eq:wave_eu1}) in cylindrical coordinates will become
   %~~~~~~~~~~~~~~~~~~~~~~~~~~~~~~~~~~~~~~~
\begin{equation}\label{eq:tsw_anal}
{\frac{\partial^2{\xi_z(\vec{r},t)}}{\partial t^2}}
 = {v_p^2} \left({\frac{\partial^2{\xi_z(\vec{r},t)}}{\partial r^2}}
+{\frac{1}{r}}{\frac{\partial{\xi_z(\vec{r},t)}}{\partial r}}\right){.}  
  \end{equation} 
  %~~~~~~~~~~~~~~~~~~~~~~~~~~~~~~~~~~~~~~~
Here, $r= (x^2+y^2)^{1/2}$. The solution are Bessel functions which for large r approach asymptotically\cite{sekeljic2010asymptotic} to
  %~~~~~~~~~~~~~~~~~~~~~~~~~~~~~~~~~~~~~~~
  \begin{equation}\label{}
{\xi_z}(\vec{r},t)= {\xi_z}(\vec{r},\omega){\approx}{\frac{\vec{\xi_{z0}}}{\sqrt{r}}}{e^{-j{\omega}t}}
  \end{equation}
  %~~~~~~~~~~~~~~~~~~~~~~~~~~~~~~~~~~~~~~~
  Here, angular frequency $ \omega=k\,v_p$, and $ k$ is wavenumber. The associated wavefronts are cylindrical which propagate radially outward at the phase velocity $v_p= \omega/k=\sqrt{{\eta}/{{\rho_d}{\tau_m}}}$ and their amplitude decrease as $1/{\sqrt{r}}$.  If the wavefront emerging/collapsing from/into a point is spherical, their amplitude must attenuate as $1/r$. A  planar source will produce plane wavefronts traveling with a constant amplitude $i.e.$ the wave does not attenuate. The amplitude scaling of these wavefronts  is related to the energy conservation consideration. Note that the present simulations have been carried out in two-dimensional (x-y plane) which is the plane of rotation for vorticity structures.  So, a numerically expected transverse wave should meet the conditions for the cylindrical case.
%%%%%%%%%%%%%%%%%%%%%%%%%%%%%%%%%%%%%%%
% Numerical simulation
%%%%%%%%%%%%%%%%%%%%%%%%%%%%%%%%%%%%%%%
\section{Results discussion and numerical simulation}
 \label{num_methodology}
For the numerical simulation the generalized momentum Eq.~(\ref{eq:momentum1}) has been expressed as a set of following two coupled convective equations,
  %~~~~~~~~~~~~~~~~~~~~~~~~~~~~~~~~~~~~~~~
 \begin{eqnarray}\label{eq:vort_incomp1}
{{\rho_d}\left(\frac{\partial{\vec{v}_d}}{\partial {t}}+{\vec{v}_d}{\cdot} \nabla{\vec{v}_d}\right)}+{\alpha}{\nabla}{{\rho_d}}+{\rho_c}\nabla \phi_{d} ={\vec \psi}
 \end{eqnarray}
%~~~~~~~~~~~~~~~~~~~~~~~~~~~~~~~~~~~~~~~
\begin{equation}\label{eq:psi_incomp1}
\frac{\partial {\vec \psi}} {\partial t}+\vec{v}_d \cdot \nabla{\vec \psi}=
{\frac{\eta}{\tau_m}}{\nabla^2}{\vec{v}_d }-{\frac{\vec \psi}{\tau_m}}{.}
\end{equation}
%~~~~~~~~~~~~~~~~~~~~~~~~~~~~~~~~~~~~~~~
For our considered 2D system of equations the above variables vary in x and y directions $i.e$ ${\vec \psi}(x,y)$, ${\vec v_d}(x,y)$, ${\rho_d}(x,y)$.  From Eq.~(\ref{eq:vort_incomp1}) it is clear that the quantity ${\vec \psi}(x,y)$ is the strain created in the elastic medium by the time-varying velocity fields. Let's take the curl of an  Eq.~(\ref{eq:vort_incomp1}). Here, the curl of the second (because pressure is a function of density through  equation of state, so the curl becomes zero) and third term (as charge density is constant, so the curl becomes zero) vanish. Thus, the final coupled numerical model equations of continuity  and  momentum  equations becomes
%~~~~~~~~~~~~~~~~~~~~~~~~~~~~~~~~~~~~~~~
\begin{equation}\label{eq:cont_incomp3}
 \frac{\partial \rho_d }{\partial t} +  \left(\vec{v}_d\cdot
\nabla\right)\rho_d= 0{,}
    \end{equation}
%~~~~~~~~~~~~~~~~~~~~~~~~~~~~~~~~~~~~~~~
\begin{equation}\label{eq:psi_incomp3}
\frac{\partial {\vec \psi}} {\partial t}+\left(\vec{v}_d \cdot \vec
\nabla\right)
{\vec \psi}={\frac{\eta}{\tau_m}}{\nabla^2}{\vec{v}_d }-{\frac{\vec
\psi}{\tau_m}}{,}  
\end{equation}

%~~~~~~~~~~~~~~~~~~~~~~~~~~~~~~~~~~~~~~~
\begin{equation}\label{eq:vort_incomp3} 
\frac{\partial{\xi}_z} {\partial t}+\left(\vec{v}_d \cdot \vec \nabla\right)
{{\xi}_z}={\frac{\partial}{\partial x}}\left({\frac{\psi_{y}}{\rho_d}}\right)
-{\frac{\partial}{\partial y}}\left({\frac{\psi_{x}}{\rho_d}}\right){.}   
\end{equation}
%~~~~~~~~~~~~~~~~~~~~~~~~~~~~~~~~~~~~~~
We have used the  LCPFCT method (Boris {\it et al.}\cite{boris_book}) to evolve the coupled set of Eqs. (\ref{eq:cont_incomp3}),~  (\ref{eq:psi_incomp3}) and (\ref{eq:vort_incomp3}) for various kinds of density profiles. This method is based on a finite difference scheme associate with the flux-corrected algorithm. The velocity at each time step has been updated by using the Poisson's equation ${\nabla^2}{\vec{v}_d}=-{\vec {\nabla}}{\times}{\vec \xi}$. This Poisson's equation has been solved by using the FISPACK \cite{swarztrauber1999fishpack}. It is point out that in this particular limit, there is nothing specific which suggests that the system corresponds to a strongly coupled dusty plasma. Thus, the results discussed in this paper would in general be relevant to any viscoelastic fluid and not be restricted to the SCDPs.

Numerically, in order to understand the dynamical response of density inhomogeneous SCDPs, we consider two cases of  circularly rotating fluid vortex, especially having  (A)  smooth rotating vortex and consider another which has a  (B) sharp cut-off. Boundary conditions are periodic along the x-axis while absorbing in y-direction throughout the entire simulation work.

%%%%%%%%%%%%%%%%%%%%%%%%%%%%%%%%%%%%%%%
%%%%%%%%%%%%%%%%%%%%%%%%%%%%%%%%%%%%%%%
\subsection{Smooth rotating vortex}
 \label{num_methodology_smooth}
\begin{figure}[h]
 \centering
 \includegraphics[width=1.0\textwidth]
 {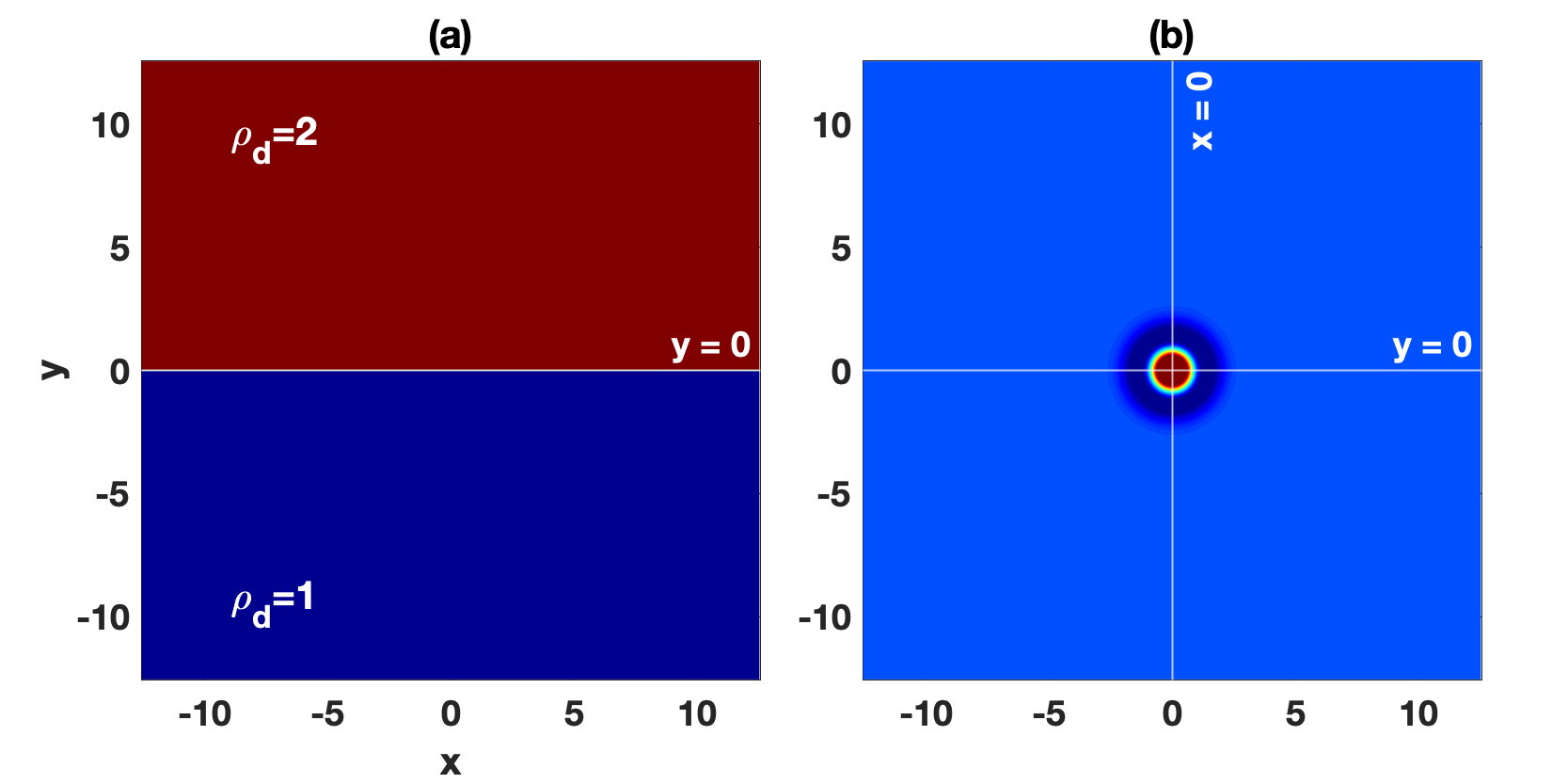}
\caption{The initial density and smooth rotating vorticity profiles. (a) Density profile where the lighter fluid is positioned side-by-side to the denser fluid  along the y=0 interface, and (b) initial profile of the smooth rotating  vorticity structure given by Eq. (\ref{eq:init_conf_vort}).}
\label{fig:vort_dens_init}	    
\end{figure}

%%%%%%%%%%%%%%%%%%%%%%%%%%%%%%%%%%%%%%%
%%%%%%%%%%%%%%%%%%%%%%%%%%%%%%%%%%%%%%%
In the former case (A), we have considered a system of length $lx=ly=8{\pi}$ units with sharp interface  of density heterogeneity shown in Fig.~\ref{fig:vort_dens_init}(a), where two incompressible fluids with different constant densities $\rho_{d}$=1 for $-{4\pi}\leq y \leq 0$ (lighter) and $\rho_{d}$=2 for $0=y \leq {4\pi}$ (denser) are positioned side-by-side along the y=0 interface. 
 %~~~~~~~~~~~~~~~~~~~~~~~~~~~~~~~~~~~~~~~
 \FloatBarrier
  %~~~~~~~~~~~~~~~~~~~~~~~~~~~~~~~~~~~~~
In interest of density waves and vorticity evolution, the sharp density interface of the medium has been perturbed through a counter-rotating vorticity structure centered at $(x,y)=(0,0)$, given by
 %~~~~~~~~~~~~~~~~~~~~~~~~~~~~~~~~~~~
  \begin{equation}\label{eq:init_conf_vort}
{\xi_{z0}(x,y,t_0)}={\Omega_0}{\left(1-{\left({x^2+y^2} \right)}\right)}
 exp\left(-\left({x^2+y^2}\right)\right){,}
\end{equation}
 %~~~~~~~~~~~~~~~~~~~~~~~~~~~~~~~~~~~
 having the velocity components which satisfy the condition of incompressibility ${\nabla}{\cdot}{\vec{v}}=0$ using ${\vec \xi}={\vec \nabla}{\times}{\vec v}$, 
${v_{x0}(x,y,t_0)}=-{\phi_0}yexp{\left(-\left({x^2+y^2}\right)\right)}$ and ${v_{y0}(x,y,t_0)}={\phi_0}xexp{\left(-\left({x^2+y^2}\right)\right)}$. Here $\Omega_0=2{\phi_0}$ is proportional to the total circulation. This vorticity profile has circular symmetry (ref. Fig.~\ref{fig:vort_dens_init}(b)).  

%%%%%%%%%%%%%%%%%%%%%%%%%%%%%%%%%%%%%%%
% Gravitational and
%%%%%%%%%%%%%%%%%%%%%%%%%%%%%%%%%%%%%%%
\subsubsection{Transverse shear waves}
\begin{figure*}[h]
\centering
 \includegraphics[width=1.0\textwidth]{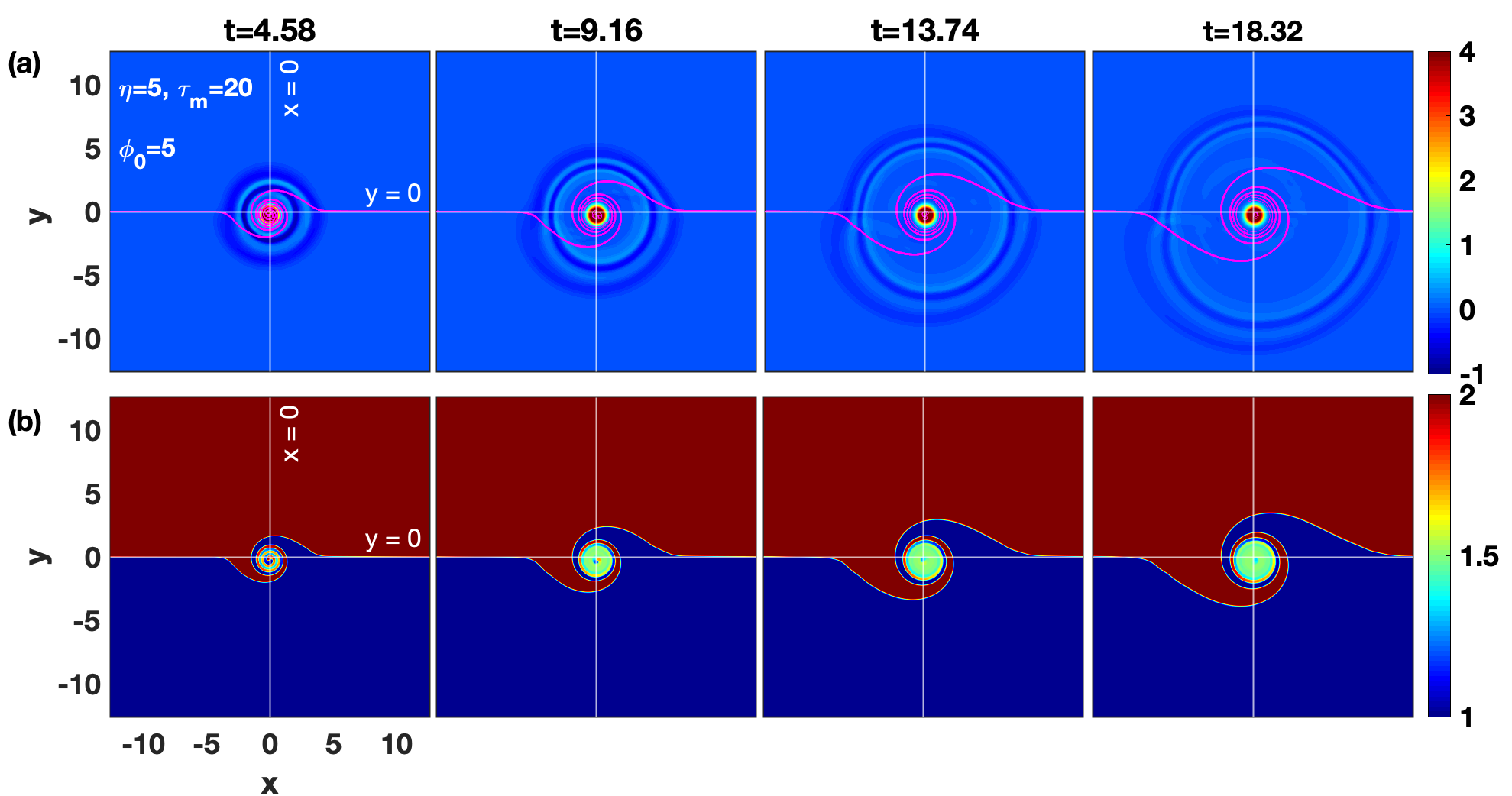}
\caption{Time evolution of a counter-rotating circular vorticity profile  (${\Omega_0}=10$) in a medium ($\eta$=5; $\tau_m$=20) where density inhomogeneity is along the interface y=0. (a) Evolution of a counter-rotating circular vorticity (Fig.~\ref{fig:vort_dens_init}(b)), where the magenta color solid line is just the interface of respective density profile; the colorbar indicates the vorticity. In (b) density profile evolution (Fig.~\ref{fig:vort_dens_init}(a)); the colorbar indicates the density.}
\label{fig:z_vort_dense_520}	    
\end{figure*}

\begin{figure}[h]
\centering
\includegraphics[width=1.0\textwidth]{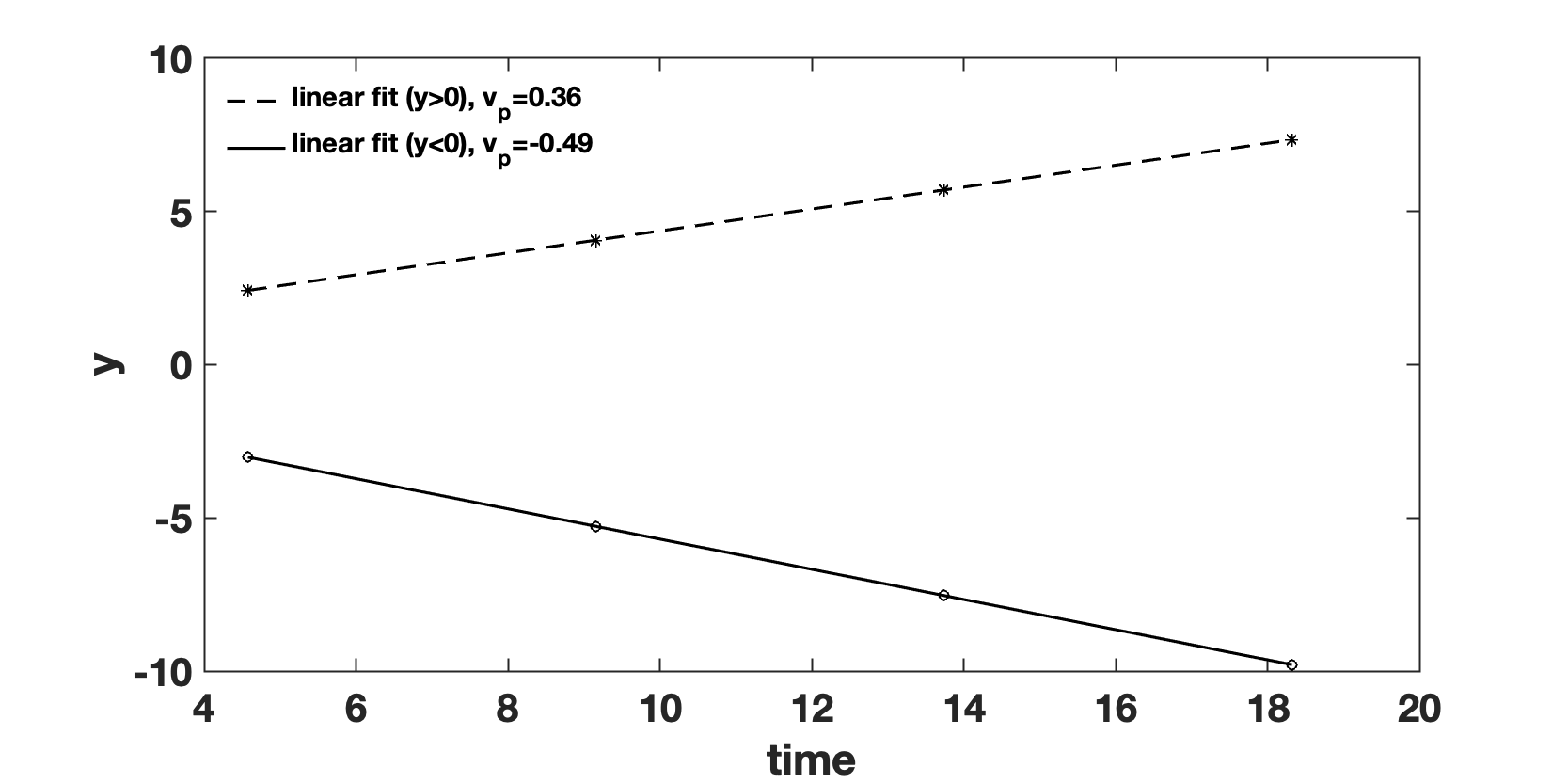}
\caption{Position-time graph  of a radially propagating wavefront (along the y-axis for x=0) observed in Fig.~\ref{fig:z_vort_dense_520}(a) for coupling parameter values $\eta$= 5; $\tau_m$=20, where $v_p$ is the phase velocity (slope of position-time plot). Here, dotted line is the linear curve fit for the denser side $(y>0)$ and solid line is the linear curve fit for the lighter side $(y<0)$. It is observed that the waves travel faster ($v_p{\approx}0.5$) in lighter side while at the slow rate ($v_p{\approx}0.35$) in the denser side of the medium which confirms that these emitted waves are TS waves.}
\label{fig:slope_520}	   
 \end{figure}

\begin{figure}[h]
\centering
\includegraphics[width=1.0\textwidth]{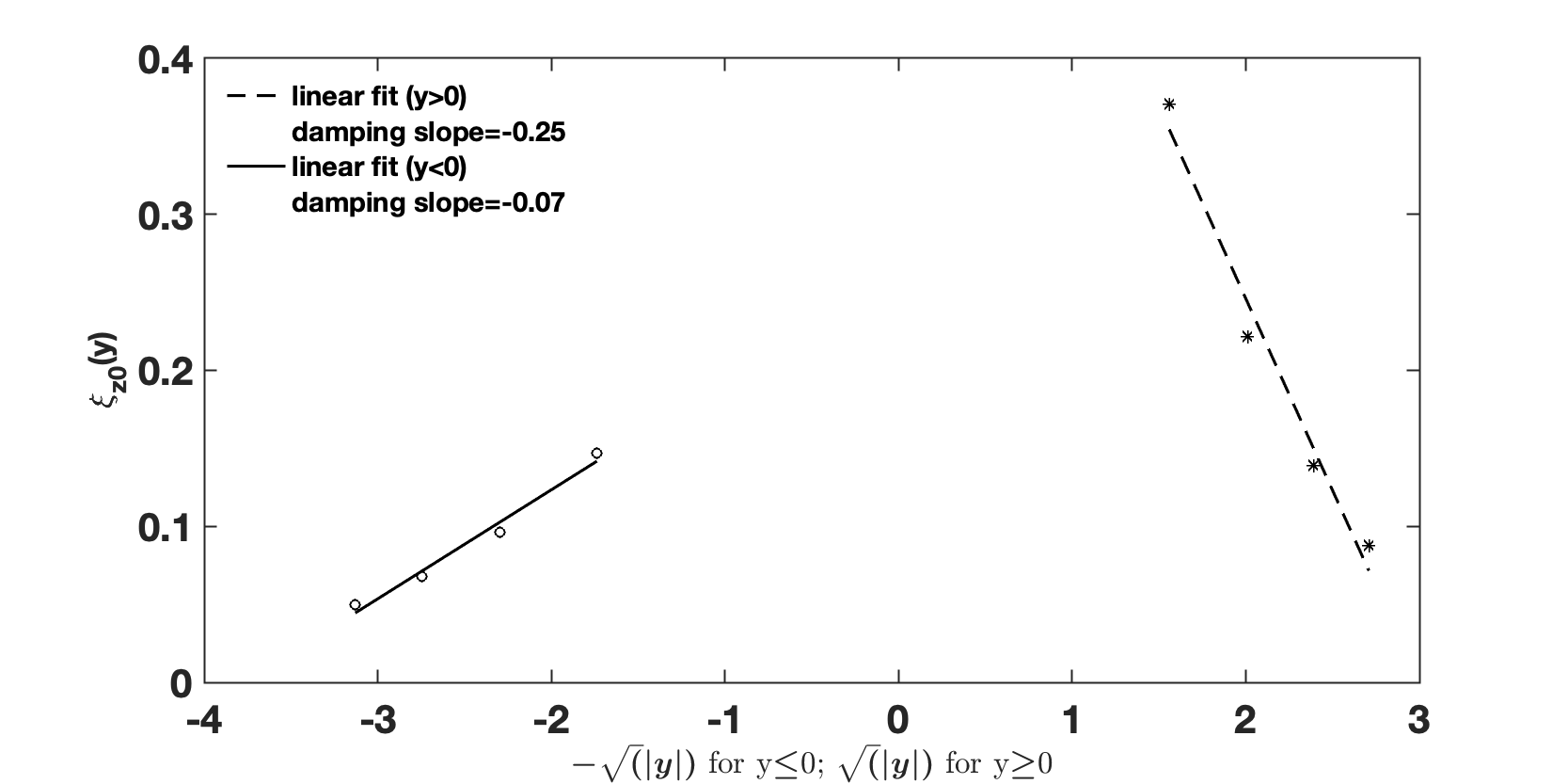}
\caption{Amplitude vs. square-root of position  of a radially propagating wavefront (along the y-axis for x=0, ref.  Fig.~\ref{fig:z_vort_dense_520}(a)) for the $\eta$= 5; $\tau_m$=20. Dotted line for the denser side $(y>0)$ is linear curve fit and solid line for the lighter side $(y<0)$ is linear curve fit.  It is observed that the amplitude of a wavefront decreases as $1/{\sqrt{y}}$ for the moving wavefront i.e. as $y$ increases. It confirms that the emitted TS waves has cylindrical shape.}
\label{fig:amp_520}	   
 \end{figure}

%\label{num_TS}
%%%%%%%%%%%%%%%%%%%%%%%%%%%%%%%%%%%%%%%
The time evolution of the above vorticity profile for ${\Omega_0}=10$ is shown in Fig.~\ref{fig:z_vort_dense_520} (a), where  $\eta=5$ and $\tau_m=20$.
   %~~~~~~~~~~~~~~~~~~~~~~~~~~~~~~~~~~~~~~~
  \FloatBarrier
  %~~~~~~~~~~~~~~~~~~~~~~~~~~~~~~~~~~~~~~~
From Fig.~\ref{fig:z_vort_dense_520}(a), as soon as the vortex starts rotating at the interface induces two things: first, the densities are rolled-up to form the spiral density waves (we will discuss it later), and second, locally introduce a shear flow. This shear flow is the source for the  radial shear waves.  These  radial waves travel a smaller distance in higher density part than in lower density part in the same time interval.  In Fig.~\ref{fig:slope_520} we have plotted the positions of a particular wavefront with time along the y-axis for $x=0$ . 
 %~~~~~~~~~~~~~~~~~~~~~~~~~~~~~~~~~~~~~~~
  \FloatBarrier
 %~~~~~~~~~~~~~~~~~~~~~~~~~~~~~~~~~~~~~~~
 In Fig.~\ref{fig:slope_520},  position-time slope suggests that the phase velocity $(v_p)$ is found to match with Eq.~(\ref{eq:TSwave_conf})  $i.e$~ $v_p{\approx}0.35$ for denser side ($\eta$= 5; $\tau_m$=20 and $\rho_d$=2 for $y>0$) and in lighter side  $v_p{\approx}5$ ($\eta$= 5; $\tau_m$=20 and $\rho_d$=1 for $y<0$). {\it {It confirms that the emerging waves are the transverse shear (TS) waves}}. Figure~\ref{fig:amp_520}  shows the radial fall of amplitude of the emitted TS wave in the  both sides (denser and lighter) of the density profile. It is found to match the fall of amplitude with $1/\sqrt{y}$ which meets the requirement of  energy  conservation for a cylindrical wave (has been discussed in the subsection~\ref{anal_description}). {\it {It confirms that the emitted TS waves have cylindrical shape}}.
   %~~~~~~~~~~~~~~~~~~~~~~~~~~~~~~~~~~~~~~~
  \FloatBarrier
 %~~~~~~~~~~~~~~~~~~~~~~~~~~~~~~~~~~~~~~~
  Hence, both the above observations (Figs.~\ref{fig:slope_520} and~\ref{fig:amp_520}) confirm that these  emerging waves are the linear cylindrical TS waves as suggested by analytical Eq.~(\ref{eq:tsw_anal}) and also indicate the validation of our numerical code.  The coupling strength parameter $\Gamma$ for a dusty plasma is proportional to the cube root of density\cite{ikezi1986coulomb} which means denser side has larger coupling strength than lighter one.  This effect can be noticed from Fig.~\ref{fig:z_vort_dense_520}(a) where the TS waves get more clearer/steeper in denser side than lighter one. Furthermore, due the density gradient and the higher speed of TS waves in lighter side than denser, the counter rotating vorticity center gets little shifted towards the low density side and the emerging TS waves are no longer cylindrical symmetric around the center of rotation. 
%%%%%%%%%%%%%%%%%%%%%%%%%%%%%%%%%%%%%%%
%%%%%%%%%%%%%%%%%%%%%%%%%%%%%%%%%%%%%%%
\subsubsection{Formation and evolution of spiral density waves}
\begin{figure*}[h]
  	\centering
  	\includegraphics[width=1.0\textwidth]{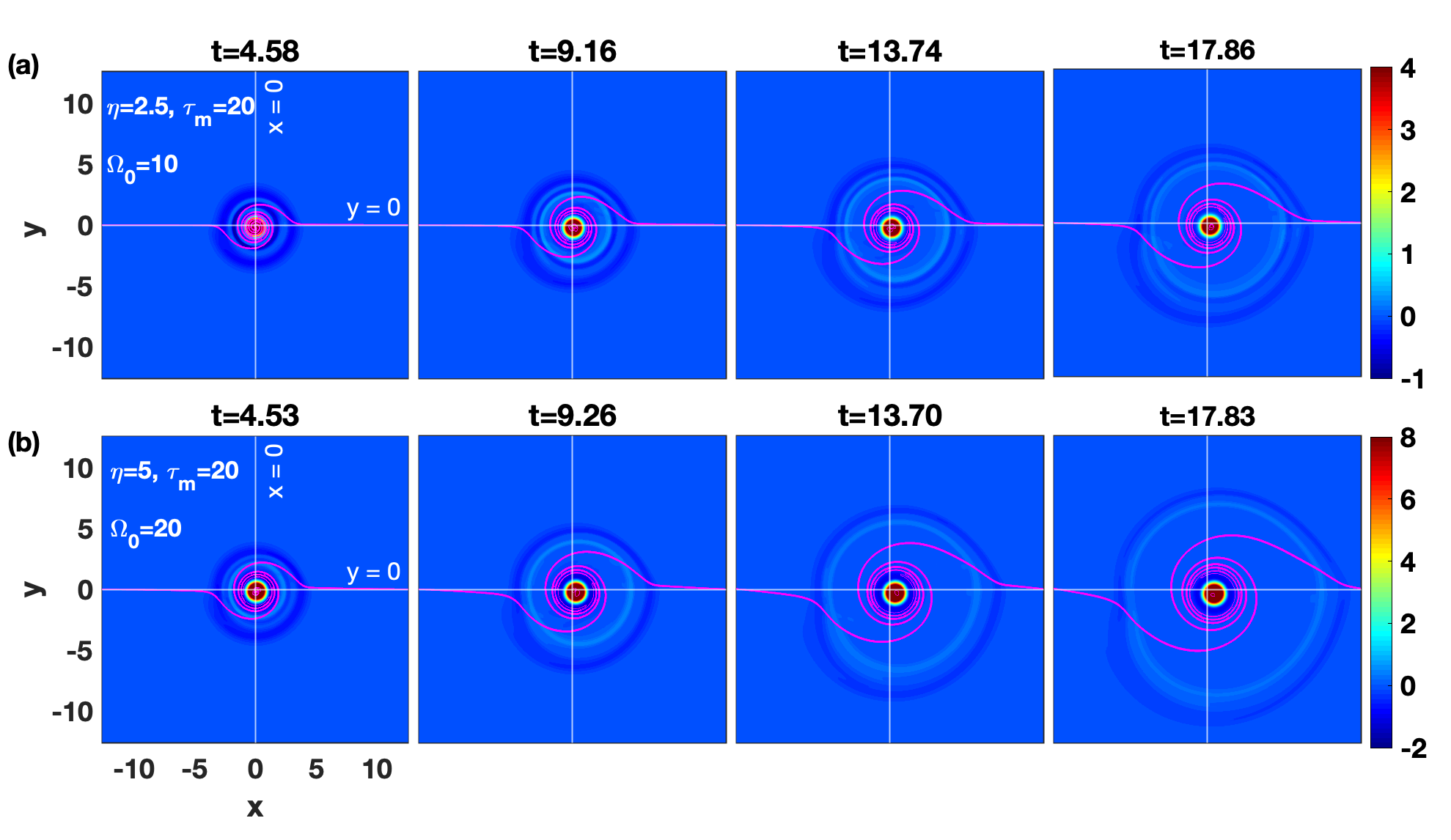}
  	\caption{Time evolution of a rotating circular vorticity profile in a medium of sharp density inhomogeneity  along the interface y=0, where the line plot (magenta color) represents the density interface over the respective rotating vorticity. (a) ${\Omega_0}=10$ and $\eta$= 2.5; $\tau_m$=20, and  (b) ${\Omega_0}=20$ and $\eta$= 5; $\tau_m$=20.  It is observed that the perturbed density region along the interface follows the radially propagating TS waves and remains confined within the distance traveled by the TS waves at that moment.  Conclusion, the impact of spiral waves can be controlled by tweaking the coupling parameter values.}
  	\label{fig:z_vort_2p5_520_om10_20}	   
  \end{figure*}

\begin{figure*}[h]
 	\centering
 	\includegraphics[width=1.0\textwidth]{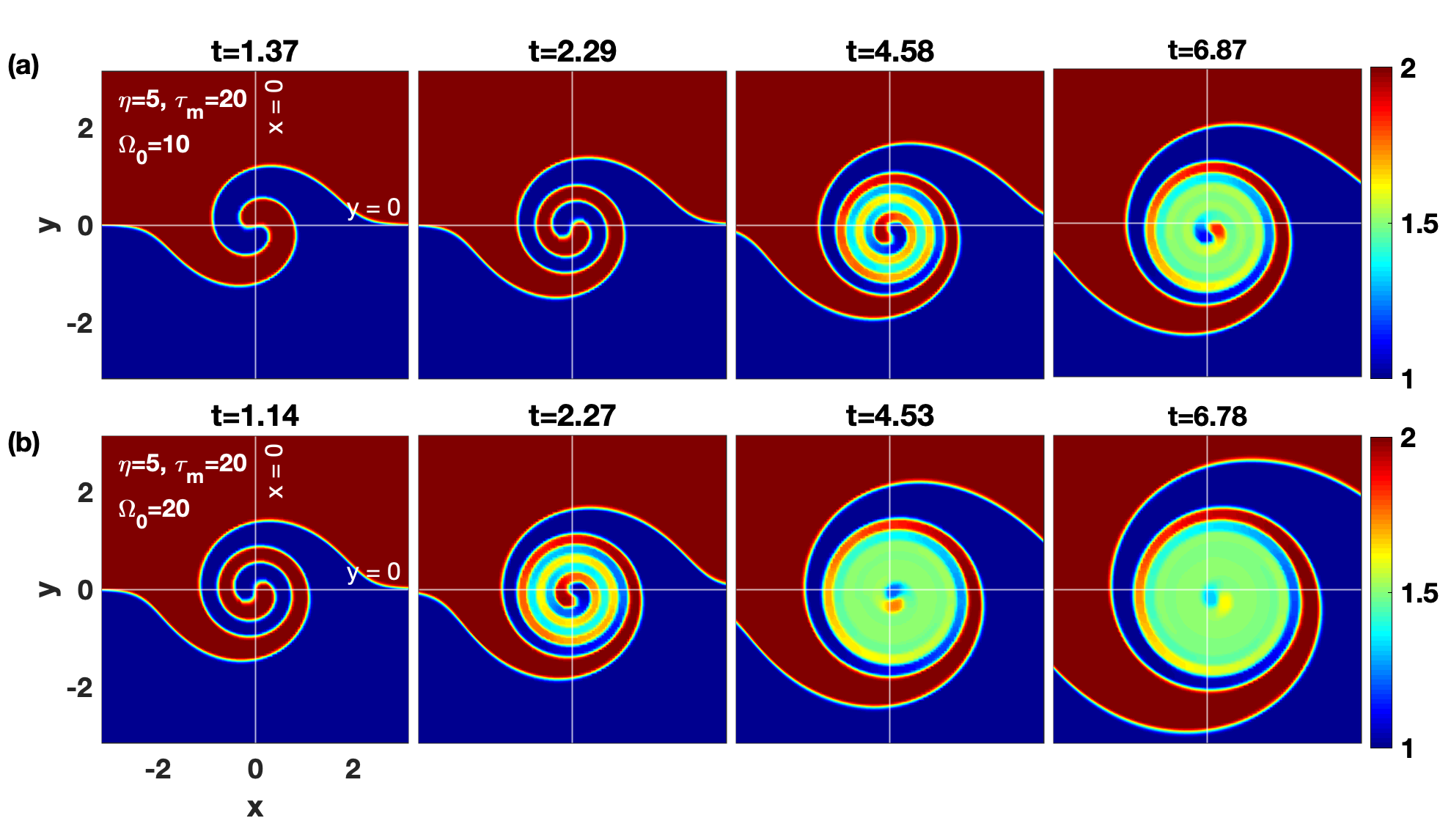}
 	\caption{Time evolution of sharp density profile for strongly coupled dusty plasma medium, where $\eta$= 5; $\tau_m$=20. (a) ${\Omega_0}=10$ and (b) ${\Omega_0}=20$.  The expansion rate (radial velocity) is high and the number of turns (angular velocity) of the spiral arms are almost double in (b) in comparison to (a) due to the two times of the angular velocity. Furthermore, the spiral arms of both densities (lighter and denser) are distinguishable in the early time while later get smeared out. }
 	\label{fig:z_spiral_dense_both}	   
 \end{figure*}

\begin{figure*}[h]
\centering
\includegraphics[width=1.0\textwidth]{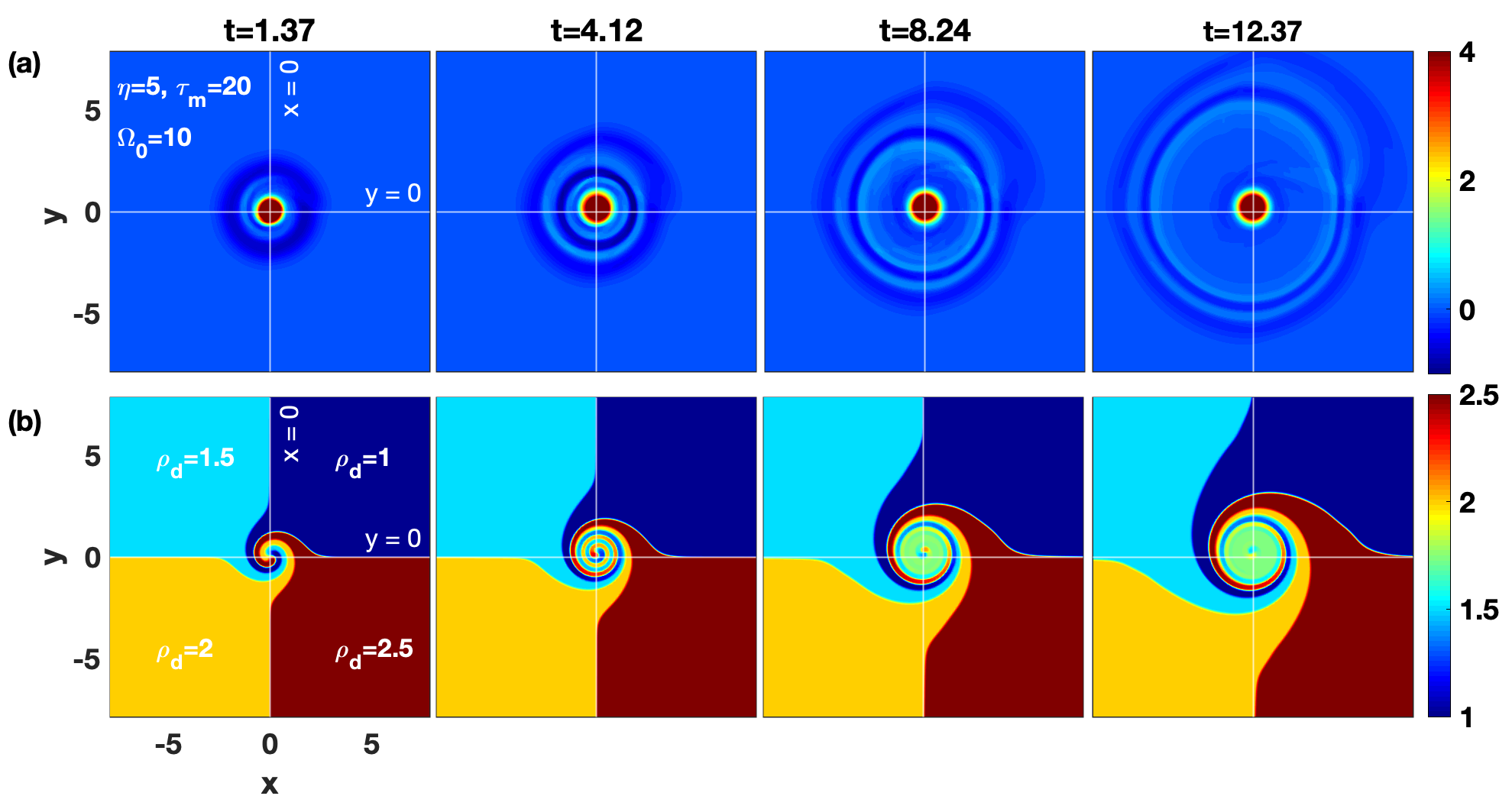}
\caption{Time evolution of four different sharp densities profiles over the rotating coherent structure for SCDPs for the $\eta$= 5; $\tau_m$=20. In (a)  colorbar designates the vorticity.  It can be noticed: The greater the density/$\Gamma$ of the medium, the steeper the TS waves; the lighter the density of a medium, the faster the speed of TS waves; the speed difference in different quadrants and the density gradient cause the wavefronts are no longer cylindrical symmetric around the center of rotaion and the shift in the vorticity center towards the low density quadrant and, in (b) four distinguishable spiral arms for each of density is observed in the early time while later get smeared out; colorbar shows the density.}
\label{fig:z_vort_dense_520_mutlti_armed.png}	   
 \end{figure*}

%\label{num_TS}
%%%%%%%%%%%%%%%%%%%%%%%%%%%%%%%%%%%%%%%
 After the confirmation of TS waves, let's focus on the evolution of background density profile shown in Fig.~\ref{fig:z_vort_dense_520}(b), where the rotating vortex basically convects material from the high denser part to the less dense  part and vice versa which results the spiral density arms/waves around the vortex  with time. One of the main advantages of viscoelastic fluids is that a rotating vortex in such media can drive density waves for a long-time duration without much dissipation  than in hydrodynamic fluids. A comparative analysis of evolution of vorticity structures in  Fig.~\ref{fig:z_vort_dense_520}(a)  to the density spiral structures in  Fig.~\ref{fig:z_vort_dense_520}(b) shows that the perturbed density region along the interface follows the radially propagating TS waves and remains confined within the distance traveled by the TS waves at that moment. To make it more clear we have also plotted the interface of respective density profile over the vorticity using a magenta colored solid line in  Fig.~\ref{fig:z_vort_dense_520}(a). In order to substantiate our observations, we have also simulated two more cases with same background density inhomogeneity, first (Fig.~\ref{fig:z_vort_2p5_520_om10_20}(a)) with same rotation rate of vorticity (${\Omega_0}=10$) but with different values of coupling parameters ($\eta$= 2.5; $\tau_m$=20)  i.e. TS waves travel with less phase velocity and second (Fig.~\ref{fig:z_vort_2p5_520_om10_20}(b)) with double rotation rate of vorticity (${\Omega_0}=20$) but with same values of coupling parameters ($\eta$= 5; $\tau_m$=20) i.e. TS waves travel with same phase velocity (Fig.~\ref{fig:z_vort_dense_520}(a)). It can be seen that these two cases too favor our earlier observation and implies that the spiral waves can be controlled by tweaking the coupling parameter values. 
 %  Moreover, the effect of high rotation rate is observed in form of bit larger vorticity center displacement and large expansion of spiral-arms in Fig.~\ref{fig:z_vort_2p5_520_om10_20}(b) than Fig.~\ref{fig:z_vort_dense_520}(a).
   %~~~~~~~~~~~~~~~~~~~~~~~~~~~~~~~~~~~~~~~
  \FloatBarrier
  %~~~~~~~~~~~~~~~~~~~~~~~~~~~~~~~~~~~~~~~
These observations can be visualised more clearly through the zoomed density contour snapshots shown in Fig.~\ref{fig:z_spiral_dense_both}.
 %~~~~~~~~~~~~~~~~~~~~~~~~~~~~~~~~~~~~~
 \FloatBarrier
%~~~~~~~~~~~~~~~~~~~~~~~~~~~~~~~~~~~~~
Spiral waves have both angular as well as radial velocity components. A comparison between Fig.~\ref{fig:z_spiral_dense_both}(a) ($\Omega_0=10$) and Fig.~\ref{fig:z_spiral_dense_both}(b) ($\Omega_0=20$) shows that the expansion rate (radial velocity) is high and the number of turns (angular velocity) of the spiral arms are almost double in Fig.~\ref{fig:z_spiral_dense_both}(b) in comparison to Fig.~\ref{fig:z_spiral_dense_both}(a) due to the two times of the angular velocity. Thus, the conclusion is that the number of turns of spiral arms are proportional to the amplitude of  vorticity. Furthermore, the spiral arms of both densities (lighter and denser) are distinguishable in the early time while later get smeared out.  So far, we have dealt with a medium having two fluids of different densities which results in the formation of two kind of spiral arms. Next we have simulated a medium has four fluids of different densities over the rotating coherent structure as shown in Figure~\ref{fig:z_vort_dense_520_mutlti_armed.png}. The boundary conditions have been taken care of by considering all the four density quadrants are localised and lie inside the density region considered in Fig.~\ref{fig:vort_dens_init}(a). 
%~~~~~~~~~~~~~~~~~~~~~~~~~~~~~~~~~~~~~
 \FloatBarrier
%~~~~~~~~~~~~~~~~~~~~~~~~~~~~~~~~~~~~~~~
 Again from Fig.~\ref{fig:z_vort_dense_520_mutlti_armed.png} it can be noticed: The greater the density/$\Gamma$ of the medium, the steeper the TS waves; the lighter the density of a medium, the faster the speed of TS waves; the speed difference in different quadrants and the density gradient cause the wavefronts are no longer cylindrical symmetric around the center of rotaionans and the shift in the vorticity center towards the low density quadrant.  In Fig.~\ref{fig:z_vort_dense_520_mutlti_armed.png}(b) four distinguishable spiral arms for each of density is observed in the early time while later get smeared out.  Thus,  from all the above evolutions of the density profiles it can be anticipated that the number spiral arms are proportional to the number of different densities coexist over the smooth rotating structure.
%%%%%%%%%%%%%%%%%%%%%%%%%%%%%%%%%%%%%%%
%%%%%%%%%%%%%%%%%%%%%%%%%%%%%%%%%%%%%%%
\subsection{Rigid rotating vortex}
 \label{num_methodology_rigid}
\begin{figure*}[h]
	\centering
	\includegraphics[width=1.0\textwidth]{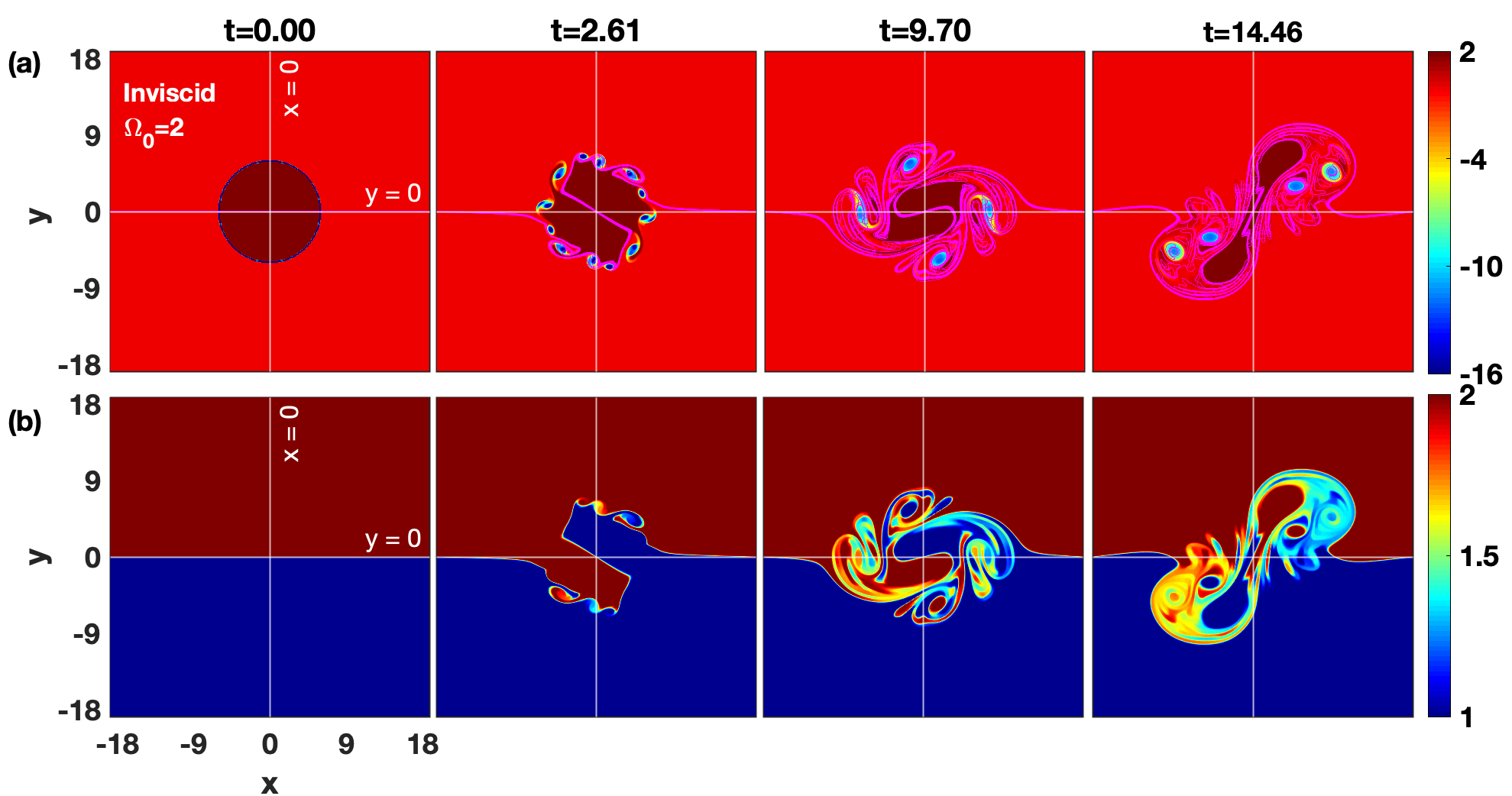}
\caption{Time evolution of a sharp counter-rotating vorticity profile  (${\Omega_0}=10$) in a inviscid HD fluid ($\eta$=5; $\tau_m$=20) with background density inhomogeneity along the interface y=0. (a) Evolution of a sharp counter-rotating vorticity, where the magenta color solid line is just the interface of respective density profile and (b) density profile evolution}
	\label{fig:z_sharp_vort_dense_inviscid}	   
\end{figure*}

\begin{figure*}[h]
 	\centering
 	\includegraphics[width=1.0\textwidth]{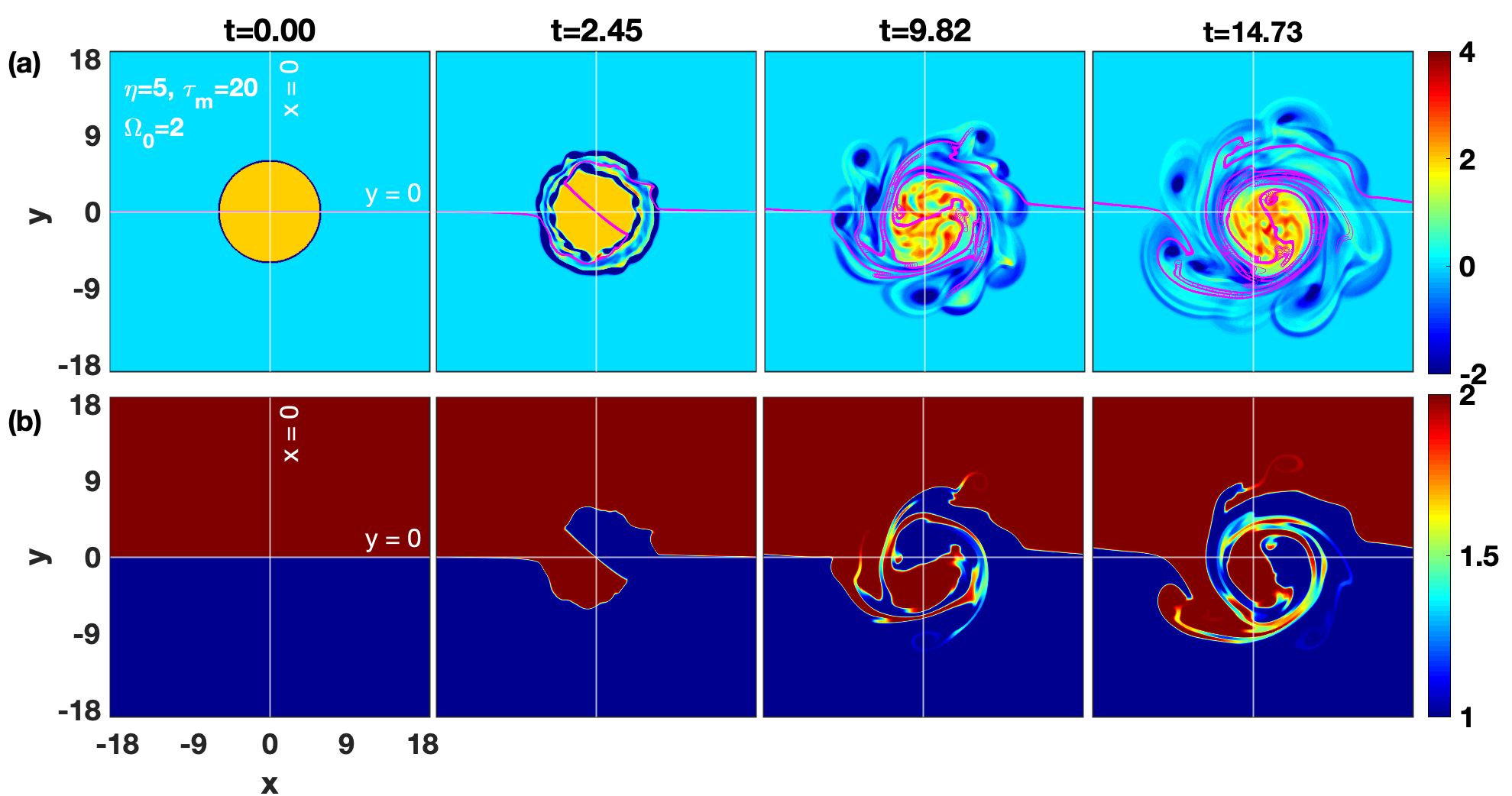}
\caption{Time evolution of a sharp counter-rotating vorticity for GHD system where $\eta$=5; $\tau_m$=20 with sharp background density inhomogeneity along the interface y=0. (a) Evolution of a sharp  vorticity is ${\xi}_{z0} = 0$ beyond $r = r_0$ ($ =6.0$) and  for $r \leq r_0$ the vorticity is  taken to have a  constant value ${{\xi}_{z0}}=2{\phi_0}$; ${\phi_0}$=1 , where the magenta color solid line is just the interface of respective density profile and (b) density profile evolution.}
 	\label{fig:z_sharp_vort_dense_520}	   
 \end{figure*}

%%%%%%%%%%%%%%%%%%%%%%%%%%%%%%%%%%%%%%%
Thus far, we had specifically avoided the development of fluid Kelvin-Helmholtz (KH) instability by taking smooth flow profiles. Here, we take vorticity profile B with a sharp cut-off (first snapshot at $t=0$ in Figs.~\ref{fig:z_sharp_vort_dense_inviscid}(a) and ~\ref{fig:z_sharp_vort_dense_520}(a)) i.e. setting the vorticity ${\xi}_{z0} = 0$ beyond $r = r_0$ ($ =6.0$) and  for $r \leq r_0$ the vorticity is  taken to have a  constant value ${{\xi}_{z0}}=2{\phi_0}$; ${\phi_0}$=1.  Another possibility which we explored in our previous paper~\cite{dharodi2014visco} was the  evolution of the same sharp cut-off rotating vortex in the SCDPs  with constant background density. In the present paper, we introduce an additional density inhomogeneity where  $\rho_{d}$=1 (lighter) for $-{6\pi}\leq y \leq 0$ and $\rho_{d}$=2 (denser) for $0=y \leq {6\pi}$ (first snapshot at $t=0$ in Figs.~\ref{fig:z_sharp_vort_dense_inviscid}(b) and ~\ref{fig:z_sharp_vort_dense_520}(b)). For the HD system, it is evident from the Fig.~\ref{fig:z_sharp_vort_dense_inviscid}  that turning on the vortex the abruptness of vorticity profile generates a strong rotational sheared flow, which in turn is drastically unstable to the KH instability leads to the creation of small KH vortices across the vorticity interface and after that as time progresses the merging of these vortices takes place. This creation and merging process of  KH vortices distorts the formation of the spiral waves around the rotor. 
%~~~~~~~~~~~~~~~~~~~~~~~~~~~~~~~~~~~~~
\FloatBarrier
%~~~~~~~~~~~~~~~~~~~~~~~~~~~~~~~~~~~~~
We know that an incompressible GHD system, besides KH instability,  also favours  the TS waves emission. In  Fig.~\ref{fig:z_sharp_vort_dense_520}(a)  where $\eta$= 5; $\tau_m$=20,  we observe  a pair of ingoing and outgoing wavefronts emanates from the interface and a concomitant KH destabilization at each of these fronts.  
 %~~~~~~~~~~~~~~~~~~~~~~~~~~~~~~~~~~~~~
 \FloatBarrier
 %~~~~~~~~~~~~~~~~~~~~~~~~~~~~~~~~~~~~~
 The response of the respective background fluid density to the rigid rotor (Fig.~\ref{fig:z_sharp_vort_dense_520}(a))  is shown in Fig.~\ref{fig:z_sharp_vort_dense_520}(b) where as soon as the rotor begins to rotate: initially, the fluid within the inner region  $(|r| \le 6)$ starts rotating with it which later on undergoes mixing due to the ingoing wave from the interface while the stagnant fluid in the outer region $(|r| \ge 6)$ undergoes mixing due to the outgoing wave from the interface, and the fluid near the interface starts binding around it. The  interplay between the ingoing and outgoing wavefronts, and KH instability distorts the formation of spiral waves around the rotor. But here, the TS waves are help in efficient mixing of the viscoelastic fluids by convecting the fluids inside and outside the vortex structure. Additionally, as the TS waves travel faster in lower density part  than upper density part and the density gradient cause the counter rotating vorticity center gets shifted towards the low density side while a HD fluid does not support any TS wave so there is no shifting in the center of rotaion.  Here, it should be noted that the HD results are used only to facilitate the understanding of our observations of GHD system not for any comparative analysis.
%%%%%%%%%%%%%%%
% CONCLUSIONS %
%%%%%%%%%%%%%%%
\section{Conclusions and Outlook}
\label{Conclusions}
Dusty plasmas support the long-lived coherent structures which can provide a driving force for the generation of different types of waves and instabilities.  Here, we have examined how the density heterogeneity in dusty plasmas responds to the circularly rotating vortex monopoles, namely smooth and sharp cut-off. We have carried out the 2D fluid simulations in the framework of generalized hydrodynamics fluid model. A fluid description is applicable if  all  spacial  scales  being  considered are about an order of magnitude larger that the interparticle distance. The rotating vortices are placed at the interface of different densities. The smooth rotating vortex causes to  form the spiral density waves and radial emission of TS waves into the surrounding media according to the shear wave speed.  It is noticed that the spiral density waves are distinguishable in the early time while later get smeared out. Spiral waves are observed in many biological, physical, and chemical systems\cite{allessie1976circus,gerisch1965stadienspezifische,bodenschatz1991transitions,net1995binary,agladze2000waves}.  A sharp rotating vortex favors the Kelvin-Helmholtz (KH) instability across its interface. In such flows for the GHD system, the interplay between the emitted TS waves and the vortices of KH instability distorts the formation of the regular spiral density arms around the rotor. It  is observed that the TS waves  play an important role in controlling the transport properties like mixing, and diffusion. Although, here, the density inhomogeneity has been introduced hypothetically (in order to identify the TS waves) for simulation but the presented results can be easily generalized to any typical SCDP experiment having realistic density inhomogeneity $e.g.$, sech type, parabolic type, etc. and by locating the driving vortex at the required place.  In 3D system the TS waves may be less effective because the amplitude of a emerging TS wave from a spherical source will decrease faster ($1/r$) than the present 2D case (here, $1/\sqrt{r}$). Also, here, we have only explored a rotating coherent vorticity structures specifically, smooth and sharp cut-off. It would also be interesting to see the evolution of such inhomogeneous media having other types of coherent structures like elliptical, dipole, and tripole.
  \FloatBarrier
%%%%%%%%%%%%%%%%%%%%%%%%%%%%%%%%%%%%
\bibliographystyle{unsrt}
 \newpage
%\bibliography{vikram_spiral}

%~~~~~~~~~~~~~~~~~~~~~~~~~
\end{document}